\documentclass[preprint,12pt,useAMS]{elsarticle}

\journal{Physics of the Dark Universe}

\begin{document}

\begin{frontmatter}



\title{Two mysterious universal dark matter-baryon relations in galaxies and galaxy clusters}


\author{Man Ho Chan}

\address{Department of Science and Environmental Studies, The Education University of Hong Kong \\ 
Tai Po, New Territories, Hong Kong, China}
 
\begin{abstract}
Starting from the 1970s, some relations connecting dark matter and baryons were discovered, such as the Tully-Fisher relation. However, many of the relations found in galaxies are quite different from that found in galaxy clusters. Here, we report two new mysterious universal relations connecting dark matter and baryons in both galaxies and galaxy clusters. The first relation indicates that the total dynamical mass of a galaxy or a galaxy cluster $M_{500}$ has a power-law relation with its total baryonic mass $M_b$ within the `virial region': $M_{500} \propto M_b^a$, with $a \approx 3/4$. The second relation indicates that the enclosed dynamical mass $M_d$ is almost directly proportional to the baryonic mass for galaxies and galaxy clusters within the central baryonic region: $M_d \propto M_b$. The close relations between dark matter and baryons in both galaxies and galaxy clusters suggest that some unknown interaction or interplay except gravity might exist between dark matter and baryons.
\end{abstract}

\begin{keyword}
Dark matter
\end{keyword}

\end{frontmatter}



\section{Introduction}
The Tully-Fisher relation in spiral galaxies \cite{Tully} and the Faber-Jackson relation in elliptical galaxies \cite{Faber} are the earliest relations discovered that connect the dynamical property (the flat rotation velocity $V$ or velocity dispersion $\sigma$) of galaxies with the observed galactic luminosity (mainly contributed by baryons). Since then, many studies have started to examine the explicit relations between dark matter and baryons in galaxies, such as the mass-discrepancy-acceleration relation \cite{McGaugh} and the radial acceleration relation \cite{McGaugh2}. Not only in galaxies, some early studies have also discovered a correlation between the hot gas mass and hot gas temperature $M_{\rm gas} \propto T^2$ \cite{Sanders2,Mohr} and the virial mass-temperature relation \cite{Ventimiglia,Gonzalez} in galaxy clusters. However, many of the relations found in galaxies are quite different from that in galaxy clusters. For example, the acceleration scale found in the radial acceleration relation in galaxies are largely different from that in galaxy clusters \cite{Chan,Pradyumna,Chan3}.

Nevertheless, some recent studies start to realise that there are some universal properties for both galaxies and galaxy clusters. For example, there exists a physical constant related to the properties of dark matter and baryonic matter, in which the values for galaxies and galaxy clusters are very close to each other \cite{Chan2}. Moreover, a more recent study has discovered a tight mass-velocity dispersion relation $M_b \propto \sigma^{4.1 \pm 0.5}$ \cite{Tian}, which is analogous to the baryonic Faber-Jackson relation found in galaxies. These results not only suggest a possible universal relation connecting dark matter and baryons, but also potentially suggest the existence of some unknown interplay except gravity between dark matter and baryons.

Investigating this issue is very important because the existence of any universal relation connecting dark matter and baryons in different structures is not obvious for the cold dark matter (CDM) model. It is because the CDM model predicts that the interaction between dark matter and baryonic matter is nearly negligible. Therefore, some studies have suggested that the existence of the close relation between dark matter and baryons is better interpreted as a possible sign of modified gravity or Modified Newtonian Dynamics (MOND) \cite{Milgrom,Green}. However, some numerical studies seem to be able to reproduce some relations (e.g. the radial acceleration relation) connecting dark matter and baryons \cite{Ludlow,Stone}. These studies have initiated some debate and discussions in this particular problem \cite{Rodrigues,McGaugh3,Rodrigues2,Brouwer}. 

\section{Method} 
\subsection{The baryonic mass and dynamical mass for the SPARC galaxies}
The last data point $(r_b,V_b,V_c)$ in the Spitzer Photometry \& Accurate Rotation Curves (SPARC) database can define the enclosed baryonic mass and the enclosed dynamical mass within the outermost radius $r=r_b$ for each galaxy, where $V_b$ and $V_c$ are the rotation velocity contributed by the baryonic matter (including bulge, disk and gas) and the total rotation velocity respectively. The enclosed dynamical mass at $r=r_b$ can be easily calculated by $M_d=r_bV_c^2/G$. The enclosed baryonic mass is given by \cite{Lelli}
\begin{equation}
M_b=\frac{r_b}{G}\left[|V_{\rm gas}|V_{\rm gas}+\Upsilon_{\rm disk}|V_{\rm disk}|V_{\rm disk}+\Upsilon_{\rm bul}|V_{\rm bul}|V_{\rm bul} \right],
\end{equation}
where $\Upsilon_{\rm disk}$ and $\Upsilon_{\rm bul}$ are the stellar mass-to-light ratios (in $M_{\odot}/L_{\odot}$) for the disk and bulge components respectively. Note that the absolute signs exist in Eq.~(1) because the velocities of the baryonic components can be negative in general. The negative value means that the baryonic distribution has a significant central depression and the material in the outer regions exerts a stronger gravitational force than that in the inner parts \cite{Lelli}. Here, we have taken the benchmark values $\Upsilon_{\rm disk}=0.5$ and $\Upsilon_{\rm bul}=0.7$ for all galaxies to calculate $M_b$. To get the total mass $M_{500}$, we follow the CDM framework to write $M_{500} \approx KV_c^3$ \cite{Mo}, where $K \approx 1.6 \times 10^5M_{\odot}$ km$^{-3}$ s$^{-3}$ \cite{McGaugh5}. The total dynamical mass $M_{500}$ is defined as the enclosed mass at a certain radius $r_{500}$ where the enclosed average mass density equals 500 times of the cosmological critical density. Generally speaking, the total dynamical mass $M_{500}$ is the sum of the total dark matter mass $M_{\rm DM}$ and the total baryonic mass $M_b$. The `virial region' is defined as the whole region within the radius $r_{500}$. Strictly speaking, the ratio $M_b/M_{500}$ in a galaxy does not necessarily equal to the cosmological baryonic to dark matter ratio, though some other studies have assumed that $M_b/M_{500}$ is a constant for all galaxies.

\subsection{The baryonic mass and dynamical mass for the extended HIFLUGCS galaxy clusters}
For the hot gas mass $M_{\rm gas}$ and total mass $M_{500}$ of galaxy clusters, we follow the values calculated in an X-ray study \cite{Chen} and re-scale the values by adopting the Hubble parameter $h=0.73$ used in the SPARC analysis \cite{Lelli}. In order to convert the hot gas mass to the total baryonic mass, we apply the the following empirical relation \cite{Giodini}
\begin{equation}
f_{\rm star} \equiv \frac{M_{\rm star}}{M_{\rm gas}}=(5.0 \pm 0.1) \times 10^{-2} \left(\frac{M_{\rm gas}}{5\times 10^{13}M_{\odot}} \right)^{-0.37 \pm 0.04}.
\end{equation}
The total baryonic mass can be calculated by $M_b=M_{\rm star}+M_{\rm gas}=(f_{\rm star}+1)M_{\rm gas}$.

Moreover, we follow the X-ray studies of the hot gas in galaxy clusters to calculate the enclosed baryonic mass and dynamical mass at different $r$. X-ray observations indicate that most of the hot gas number density profiles in galaxy clusters can be well-fitted by the following $\beta$-model \cite{Reiprich,Chen}
\begin{equation}
n(r)=n_0\left(1+\frac{r^2}{r_c^2} \right)^{-3\beta/2},
\end{equation}
where $n_0$ is the central hot gas number density, $r_c$ is the core radius and $\beta$ is the profile index. All of the hot gas profiles in galaxy clusters can be empirically determined by these three fitted parameters $n_0$, $r_c$ and $\beta$. Assume that the hot gas in galaxy clusters is in hydrostatic equilibrium, the total dynamical mass profile $M(r)$ (dark matter plus baryonic matter) can be probed from the $\beta$-model \cite{Chen}: 
\begin{equation}
\frac{d}{dr}[n(r)kT]=-\frac{GM(r)\rho_g}{r^2},
\end{equation}
where $\rho_g$ is the hot gas mass density. Since the temperature of the hot gas in many galaxy clusters is almost constant \cite{Vikhlinin,Reiprich2}, we simply use the average hot gas halo temperature to represent $T$ \cite{Chen} and get the total dynamical mass density profile as follow \cite{Chan2}:
\begin{equation}
\rho(r)=\frac{1}{4\pi r^2} \frac{dM(r)}{dr}= \frac{3\beta kT}{4\pi G \mu m_p} \left[\frac{3r_c^2+r^2}{(r_c^2+r^2)^2} \right],
\end{equation}
where $\mu=0.59$ is the molecular weight and $m_p$ is the proton mass. To compare the ratio of the baryonic mass to the dynamical mass, we choose the central region $r \rightarrow 0$ for comparison. The central baryonic mass density is $\rho_{b0}=m_gn_0$, where $m_g$ is the average mass of a hot gas particle. The central enclosed average dynamical mass density is $\rho_{d0}=9\beta kT/4 \pi G\mu m_p r_c^2$. It can be shown that $\rho_{d0}$ is directly proportional to $\rho_{b0}$. To compare with the results in galaxies, we can calculate the enclosed dynamical mass and the enclosed baryonic mass within an arbitrary radius $R$. By choosing the average value of $r_b \approx 20$ kpc in the SPARC sample as the arbitrary radius $R$, we can get $M_d \approx 4 \pi \rho_{d0}R^3/3$ and $M_b \approx 4 \pi \rho_{b0}R^3/3$. Therefore, we can obtain the relation between $M_d$ and $M_b$ for galaxy clusters.

\subsection{The scatter estimation}
Assume that $X=\log M_d$ and $X_i$ correspond to the data of the dynamical mass. The values predicted by the universal relations are given by $X_p$. The value of the r.m.s. scatter (the unit is dex) in the log-space calculated by $N$ data is given by
\begin{equation}
\sigma_s=\sqrt{ \frac{1}{N-1} \sum_{i=1}^N \left(X_p-X_i \right)^2}.
\end{equation}
It can be shown that the percentage scatter of a relation is approximately equal to $\ln 10 \times \sigma_s \approx (230 \sigma_s)$\%. Therefore, for the r.m.s. scatter smaller than 0.2 dex, the percentage error is approximately less than 50\%, which shows a tight relation in general.

\section{Results}
Here, we first report the mysterious relations connecting the total dynamical mass and total baryonic mass of a galaxy or a galaxy cluster. We analyse the sample of galaxies (total number = 175) in the SPARC database \cite{Lelli}. In the database, it has the rotation velocity contributed by baryons $V_b$ and the total rotation velocity $V$ (contributed by both dark matter and baryons) for different galactic radii $r$. We can derive the enclosed baryonic mass and the total enclosed mass by $V_b$ and $V$ respectively. The total baryonic mass for each galaxy can be approximately indicated by the last data point of $V_b$ at the largest radius $r=r_b$ while the final data point of $V$ (i.e. $V_c$) can give the total enclosed mass $M_{500}$ for each galaxy. However, if the value of $V_c$ is too small, the enclosed mass at $r_b$ calculated is still dominated by baryons, which gives a bad projection of the total enclosed mass $M_{500}$. This mainly occurs in very small dwarf galaxies. Therefore, to reduce the systematic uncertainty, we neglect 24 galaxies in the SPARC sample with $V_c<50$ km/s so that we only analyse 151 galaxies in total. We plot the relation between $M_{500}$ and $M_b$ in Fig.~1.

We also analyse the data of galaxy clusters by using the extended HIFLUGCS galaxy cluster sample \cite{Chen}, which has 107 galaxy clusters. X-ray studies of these galaxy clusters have obtained some empirical parameters for the hot gas \cite{Chen}. These parameters can directly probe the total hot gas mass $M_{\rm gas}$ and the total dynamical mass $M_{500}$. The values of $M_{\rm gas}$ and $M_{500}$ for the extended HIFLUGCS galaxy cluster sample have been calculated explicitly \cite{Chen}. In our study, we aim at analysing large galaxy clusters because the mass of small galaxy clusters and galaxy groups are usually dominated by the bright cluster galaxies (BCGs) at their centres. In these cases, the total enclosed mass calculated by the hot gas profiles would be inaccurate. Therefore, we mainly focus on 64 large galaxy clusters with scale radius $r_c \ge 100$ kpc in the sample. Moreover, although the hot gas mass generally dominates the baryonic mass in a large galaxy cluster, we also include the stellar mass contribution $M_{\rm star}$ to get a more accurate calculation. Therefore, we can get the total baryonic mass $M_b=M_{\rm gas}+M_{\rm star}$ for each galaxy cluster. Note that the average hot gas mass in our sample galaxy clusters is $M_{\rm gas}=4.7 \times 10^{13}M_{\odot}$. The average stellar mass to hot gas mass ratio is only 5\%. Therefore, the stellar contribution is not significant. Besides, out of the 64 galaxy clusters considered in our analysis, there are 12 cooling-flow clusters. Although 7 clusters out of them are better fitted with the double-$\beta$ model rather than the single-$\beta$ model in Eq.~(3), the overall reduced chi-square fits for these clusters are small ($<2$), except the A2142 cluster only \cite{Chen}. This means that the single-$\beta$ model is a very good approximation for most of the galaxy clusters considered in this analysis. We finally plot the relation between $M_{500}$ and $M_b$ in Fig.~1 as well. 

Surprisingly, the best fit power-law relation for galaxies and galaxy clusters are very close to each other. For galaxies, we have $\log (M_{500}/M_{\odot})=(0.74 \pm 0.02) \log(M_b/M_{\odot})+(4.10 \pm 0.23)$ (r.m.s. scatter = 0.20 dex). For galaxy clusters, we have $\log (M_{500}/M_{\odot})=(0.75 \pm 0.05) \log(M_b/M_{\odot})+(4.45 \pm 0.68)$ (r.m.s. scatter = 0.14 dex). The slopes and the normalisation constants of the log-relations are very close to each other. In Fig.~1, we also include some data points obtained from other studies \cite{Gonzalez,McGaugh5} and we can see that they are all consistent with each other. Our results indicate that there exists a universal power-law relation connecting the total dynamical mass and the total baryonic mass for both galaxies and galaxy clusters: $M_{500} \propto M_b^a$, with $a \approx 3/4$. The small scatters (r.m.s. scatter $\le 0.2$ dex) suggest that they are tight relations. If we include the 24 neglected galaxies, this would mainly change the relation in the low-mass end. The overall slope would increase slightly to about 0.8 because the total mass would be underestimated for these small galaxies. Also, the resulting scatters would be larger.

Interestingly, this universal power-law relation can directly derive the existing relations in galaxies and galaxy clusters. From the universal relation, we have $M_b \propto M_{500}^{4/3}$. Since the standard cosmological model predicts $M_{500} \propto V_c^3$ \cite{Mo} and $M_{500} \propto \sigma^3$ \cite{Evrard}, we can get $M_b \propto V_c^4$ and $M_b \propto \sigma^4$, which are the baryonic Tully-Fisher relation observed in galaxies \cite{McGaugh4} and the baryonic Faber-Jackson relation observed in both galaxies \cite{Sanders,Barat} and galaxy clusters \cite{Tian}. Moreover, since $T \propto \sigma^2$ in the hot gas of galaxy clusters, we also have $M_b \propto T^2$, which is consistent with the observed mass-temperature relation in galaxy clusters \cite{Sanders2,Mohr}. Generally speaking, the power-law relation for galaxies is consistent with the known baryonic Tully-Fisher relation or baryonic Faber-Jackson relation, and therefore it is not an entirely new relation. However, it is surprising to have a similar relation (with almost the same slope) found in galaxy clusters as well. Therefore, our results somewhat show a universal relation in different structures, which has not been discussed too much before.

The second mysterious relation is related to the ratio of the central baryonic content to the dynamical content of galaxies and galaxy clusters. For galaxies, we analyse all of the 175 SPARC galaxies and focus on the whole baryonic regions (i.e. the whole optical regions) regardless of the morphology of galaxies. We calculate the enclosed baryonic mass $M_b$ and dynamical mass $M_d$ at $r_b$. We plot $M_d$ against $M_b$ in Fig.~2 and we can see that $M_d$ is nearly directly proportional to $M_b$. The slope of the best-fit $\log M_d-\log M_b$ plot is $0.95 \pm 0.02$ (r.m.s. scatter = 0.20 dex). If we fix the slope to be 1, the best-fit relation is $M_d \approx 4.1M_b$. This proportional relation is generally consistent with the results of some previous studies \cite{Gentile}.

For galaxy clusters, we analyse the same sample of 64 large galaxy clusters. By using the empirical parameters fitted by X-ray studies, we calculate the central dynamical mass density $\rho_d$ and the central baryonic mass density $\rho_b$ for each galaxy cluster. Surprisingly, we find that $\rho_d$ is also directly proportional to $\rho_b$ for the 64 galaxy clusters. The slope of the best-fit $\log \rho_d-\log \rho_b$ relation is $1.00 \pm 0.06$ (r.m.s. scatter = 0.15 dex). For a comparison with the relation for galaxies, we transform the baryonic mass density and dynamical mass density to baryonic mass and dynamical mass respectively by integrating the densities to a radius $R$, where $R \approx 20$ kpc is the average value of $r_b$ for the SPARC galaxies. We plot $M_d$ and $M_b$ in Fig.~2 and we can see that the data points almost lie on the line $M_d \approx 8.2 M_b$. Although the proportionality constants are different for galaxies and galaxy clusters, we can see that a universal proportional relation $M_d \propto M_b$ exists in the central baryonic region for both galaxies and galaxy clusters. The small scatters (r.m.s. scatter $\le 0.2$ dex) also suggest that they are tight relations.

\section{Discussion}
There are some interesting implications for the two mysterious relations discovered. First of all, although galaxies and galaxy clusters are very different structures in our universe, the tight relations seem to reveal some universal interplays between dark matter and baryons in both central baryonic regions and the whole virial regions of galaxies and galaxy clusters. Also, the power-law index for the central baryonic regions (log-slope $\approx 1$) is different from that for the virial regions (log-slope $\approx 3/4$). These properties are not generally expected for the CDM model because it predicts that the interaction between dark matter and baryons (except gravitational interaction) should be nearly negligible. As a result, the density profiles of dark matter and baryons would be quite different so that they should not have any close relation. In fact, some previous studies have suggested that the flat rotation curves in galaxies are indeed very `special'. The transition from baryonic disks to dark matter halo domination is too smooth so that the baryons and dark matter seem to `cooperate' with each other to achieve a flat rotation curve in galaxies \cite{Battaner}. This problem is known as the `halo-disk conspiracy' \cite{Battaner}. In view of this, our results seem to echo this suggestion and reveal another kind of dark matter-baryon `cooperation'. Some unknown interplay or self-organisation processes between dark matter and baryons might be needed to explain these relations. Recent simulations and observations have started to investigate similar possible interplay between dark and luminous matter \cite{Meneghetti}. 

On the other hand, since the first mysterious relation indicates that the total dynamical mass $M_{500}$ is a power-law relation with the baryonic mass $M_b$, this implies that the baryonic fraction $f=M_b/M_{500}$ is not a constant for galaxies and galaxy clusters within the virial region. Nevertheless, the baryons considered here are mainly stars and hot gas. If the unobserved cold gas is involved, the baryonic fraction might be almost a constant for all galaxies and galaxy clusters. In fact, a recent study suggests that the ratio of total baryonic mass to total dynamical mass is almost constant for galaxy clusters if cold gas is taken into account \cite{Farahi}. Therefore, our relations shown in this study might be only applicable for hot baryons. However, if it is the case, there may exist another power-law relation that determines the ratio of hot baryons and cold baryons \cite{Farahi}.

\begin{figure}
 \includegraphics[width=140mm]{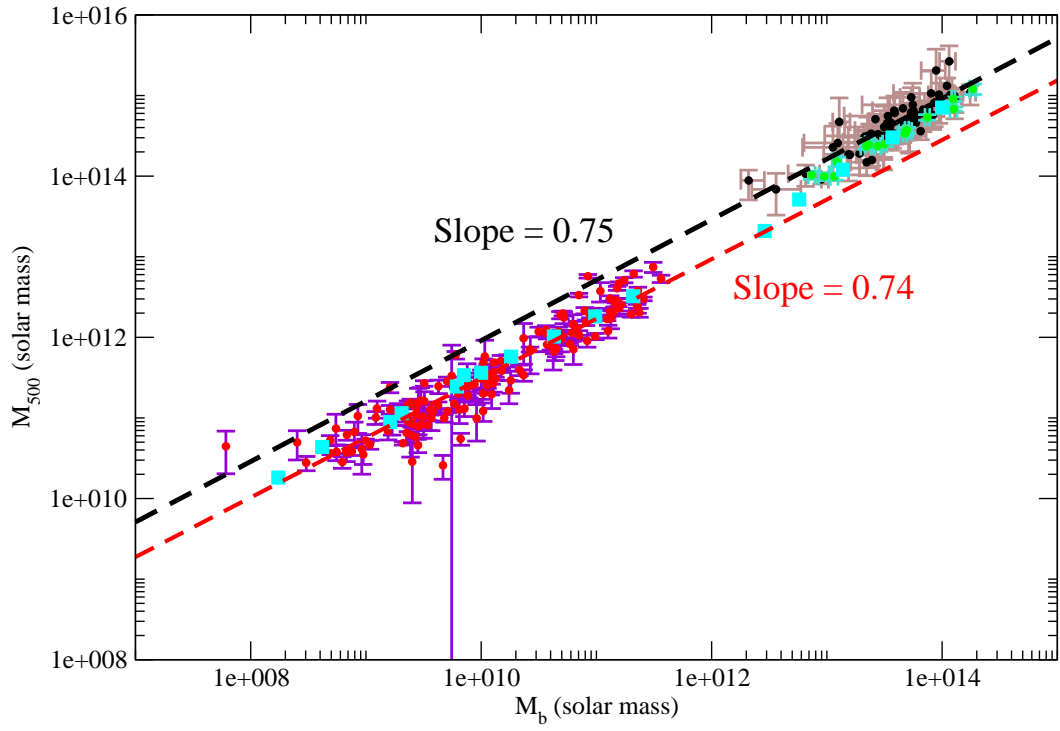}
 \caption{The relation between the total dynamical mass $M_{500}$ and total baryonic mass $M_b$. The black data points are the data of the extended HIFLUGCS galaxy cluster sample \cite{Chen}. The red data points are the data of the SPARC sample \cite{Lelli}. The green data points \cite{Gonzalez} and the blue squares \cite{McGaugh5} are the data from two other studies.}
\end{figure}

\begin{figure}
 \includegraphics[width=140mm]{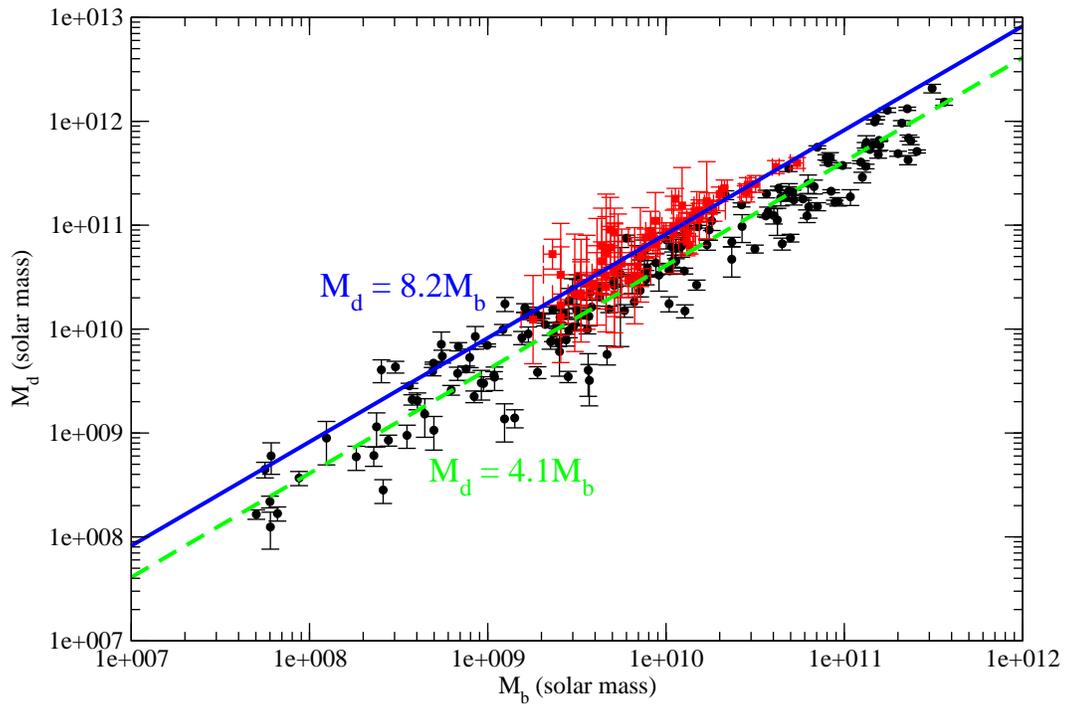}
 \caption{The relation between the enclosed dynamical mass $M_d$ and the enclosed baryonic mass $M_b$. The black and red data points represent the relations for galaxies (the SPARC sample) and the galaxy clusters (the extended HIFLUGCS sample) respectively.}
\end{figure}

\section{Acknowledgements}
The work described in this paper was partially supported by the Seed Funding Grant (RG 68/2020-2021R) and the Dean's Research Fund of the Faculty of Liberal Arts and Social Sciences, The Education University of Hong Kong, Hong Kong Special Administrative Region, China (Project No.: FLASS/DRF 04628).

\section{Conflict of interest} 
The author declares that he has no competing financial interests.





\end{document}